\documentclass{article}
\usepackage{spconf,amsmath,graphicx,hyperref}
\usepackage{cite}
\usepackage{multirow}
\usepackage{booktabs}
\usepackage{amssymb}
\usepackage[table]{xcolor}
\title{IPDnet2: an efficient and improved inter-channel phase difference estimation network for sound source localization}
\name{Yabo Wang$^{1,2}$, Bing Yang$^{2}$, Xiaofei Li$^{2,*}$\thanks{* Corresponding Author}}
\address{
  $^1$Zhejiang University, Hangzhou, China\\
  $^2$Westlake Institute for Advanced Study \& Westlake University, Hangzhou, China}
\begin{document}
\ninept
\maketitle
\begin{abstract}
IPDnet is our recently proposed real-time sound source localization network. It employs alternating full-band and narrow-band (B)LSTMs to learn the full-band correlation and narrow-band extraction of DP-IPD, respectively, which achieves superior performance. However, processing narrow-band independently incurs high computational complexity and the limited scalability of LSTM layers constrains the localization accuracy. In this work, we extend IPDnet to IPDnet2, improving both localization accuracy and efficiency. IPDnet2 adapts the oSpatialNet as the backbone to enhance spatial cues extraction and provide superior scalability. Additionally, a simple yet effective frequency-time pooling mechanism is proposed to compress frequency and time resolutions and thus reduce computational cost, and meanwhile not losing localization capability. Experimental results show that IPDnet2 achieves comparable localization performance with IPDnet while only requiring less than 2\% of its computation cost. Moreover, the proposed network achieves state-of-the-art SSL performance by scaling up the model size while still maintaining relatively low complexity.
\end{abstract}
\section{Introduction}
\label{sec:intro}

Sound source localization (SSL) estimates the positions of one or
multiple sound sources from multichannel recordings and plays a vital
role in applications such as video conferencing, human–computer
interaction, and embodied intelligence systems. These scenarios
typically impose stringent requirements on real-time performance
and computational efficiency, as they are usually deployed on resource-constrained devices\cite{GUAN2025110490}. A robust SSL method should not only ensure reliable localization accuracy but also maintain computational efficiency.

% Conventional SSL methods typically estimate spatial features such as time delay, generalized cross correlation (GCC), inter-channel phase/level difference (IPD/ILD) or relative transfer function (RTF),  and then establish the feature-to-location mapping. Although such methods can perform effectively under ideal acoustic conditions, their localization performance degrades significantly in real-world scenarios, as noise and reverberation interfere with the direct-path signal. In recent years, deep learning-based methods have been extensively studied and have achieved significant performance improvements compared to conventional methods. Among them, the full-band and narrow-band fusion network based methods demonstrate superior performance in the SSL task. Unlike mainstream methods that process only the full-band spectra to estimate source locations, these methods employ full-band layers and narrow-band layers to estimate the raw direct-path IPD (DP-IPD) information in one frequency band and capture the frequency correlations of DP-IPD, respectively. Full-band and narrow-band fusion networks have achieved outstanding localization performance. However, separately processing narrow-bands results in a large computational complexity.
Conventional SSL methods typically estimate spatial features such as time delay, generalized cross correlation (GCC), inter-channel phase/level difference (IPD/ILD)  \cite{4967888, Zhang2010ATM} or relative transfer function (RTF) \cite{Braun2015NarrowbandDE,Wang2018SemiSupervisedLW},  and then establish the feature-to-location mapping. Although such methods can perform effectively under ideal acoustic conditions, their localization performance often degrades significantly in real-world scenarios due to noise and reverberation interfering with the direct-path signal. In recent years, deep learning-based methods have been extensively studied,
these methods have achieved significant gains over conventional methods, especially in challenging conditions by learning complex patterns and subtle acoustic cues \cite{grumiaux2022survey,Jalayer_2025, DiazGuerra2020RobustSS,diaz2023permutation,WYBIS23,WYBipdnet,Xiao2024TFMambaAT}. In our previous works \cite{WYBIS23, WYBipdnet},  a full-band and narrow-band fusion network named IPDnet is proposed which enables real-time localization of multiple sources. Unlike mainstream methods that process exclusively on the full-band spectral, IPDnet employs full-band BLSTMs and narrow-band LSTMs to estimate the raw direct-path IPD (DP-IPD) information in one frequency band and capture the frequency correlations of DP-IPD, respectively. Although IPDnet demonstrates superior performance, its separate processing of narrow bands incurs high computational complexity, which limits its deployment on edge devices. Moreover, the poor scalability of the LSTM-based backbone restricts the localization performance\cite{10.5555/3618408.3619518}. The newly proposed TF-Mamba \cite{Gu2023MambaLS}, as an improvement to IPDnet \cite{Xiao2024TFMambaAT}, replaces LSTM with Mamba to improve localization accuracy, but still has a relatively high computational cost.

In this work, we improve the IPDnet\cite{WYBipdnet} in terms of both localization accuracy and computational complexity. Specifically, we adapt our previous proposed online SpatialNet (oSpatialNet)\cite{10570301} as the backbone to extract spatial features, which consists of interleaved cross-band and narrow-band modules, originally proposed for speech enhancement. In this work, the narrow-band module processes each frequency independently to learn spatial cues from individual narrow-bands, and models the temporal evolution of DP-IPDs as well. The cross-band module captures the full-band correlation of spatial cues, 
such as the approximately linear dependence of DP-IPD on frequency. 
Note that, this work is the first one to evaluate the oSpatialNet for SSL. Compared with the LSTM-based backbone in IPDnet, oSpatialNet exhibits superior scalability, enables flexible trade-offs between computational complexity and localization performance. Mel frequency compression and time compression are applied in \cite{hao2023fastfullsubnetacceleratefullband,trainmel} for reducing the computational cost of speech enhancement, in which the key is to represent speech signal at lower temporal and frequency resolutions without information loss. In this work, we explore the frequency and time compression mechanism for SSL. The temporal resolution for SSL is coarser than for speech enhancement, as SSL focuses on estimating source locations rather than reconstructing detailed spectral information. For example, we output one location estimation every 100 ms in this work, which is adequate for most applications. For frequency resolution, 
\begin{figure*}
    \centering
    \includegraphics[width=0.75\linewidth]{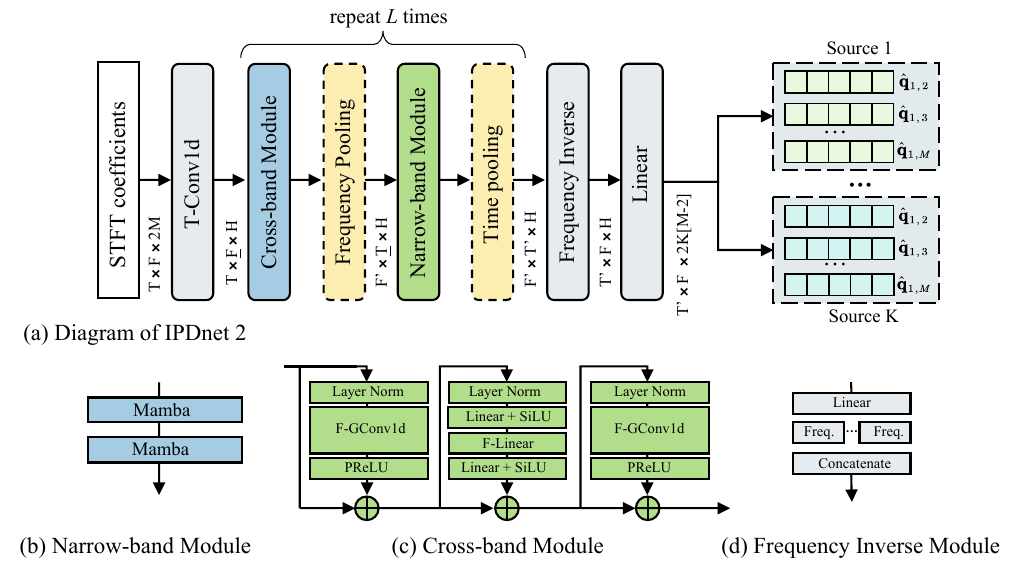}
    \caption{Model architecture of the proposed IPDnet2 network. The data organization is number of sequences $\times$ sequence length $\times$ feature dimension. Pooling layers in dashed boxes are applied only once.}
    \label{fig:overview}
\end{figure*}
\cite{hao2023fastfullsubnetacceleratefullband,trainmel} all work in the mel-frequency domain to reduce the number of frequency bands as the mel-frequency can represent speech spectra more compactly. Unlike speech enhancement, SSL does not require fine spectral fidelity, as the DP-IPDs at different frequencies intrinsically correspond to one unique time difference of arrival (TDOA) and thus requires a lower frequency resolution. Accordingly, we propose a simple yet effective frequency–time pooling mechanism and further study how the placement of frequency–time pooling compression impacts localization performance. Experimental results demonstrate that the proposed method achieves comparable localization performance with IPDnet while only requiring less than 2\% of its computation cost.  Additionally, by scaling up the model size, the proposed network achieves state-of-the-art SSL accuracy, still with relatively low computational cost.

\section{Method}
Suppose a closed environment containing multiple sound sources, along with noise and reverberation. The multichannel signals obtained from a microphone array are defined in the STFT domain as:
% \begin{equation}
% x_m(t)=\sum_{k=1}^K a_m\left(t,\theta_k\right)*s_k(t)+v_m(t),
% \end{equation}
% where $m \in[1, M]$, $k \in[1, K]$ and $t \in[1, T]$ represent the indices of microphones, sound sources and time samples, respectively. 
% % Note that in this task, $M$ is not a fixed value that we use $M$ to denote the number of microphones for presentation simplicity. 
% % $x_m(t)$, $s_k(t)$ and represent the microphone signal, source and noise signals, respectively. 
% As for the $k$-th source, $s_k(t)$, $\theta_k$, and $a_m\left(t, \theta_k\right)$ represent the source signal, the direction of arrival (DOA), and the direct-path response (within the room impulse response, RIR) to the $m$-th microphone, respectively, and * denotes convolution. The noise signal $v_m(t)$ includes both ambient noise and the reflections/reverberation of sources.

% Applying the short-time Fourier transform (STFT), the multichannel signals are expressed as:  
\begin{equation}
X_m(t, f)=\sum_{k=1}^K A_m\left(f, \theta_k\right) S_k(t, f)+V_m(t, f),
\end{equation}
where $m \in[1, M]$, $k \in[1, K]$, $t \in[1, T]$ and $f \in[1, F]$  represent the indices of microphone, sound source, time frame and frequency, respectively. $X_m(t, f)$, $S_k(t, f)$ and $V_m(t, f)$ are the STFT coefficients of microphone, source and noise signals, respectively. Note that all reflections and reverberation components are included in the noise signals. $\theta_k$ represents the direction of arrival (DOA). $A_m\left(f, \theta_k\right)$ is the transfer function of the direct-path response. 
% The room transfer function consists of the direct-path and the reflection-path component.
% \begin{equation}
% A_m(n,f, \theta)=A_m^{\mathrm{d}}(n,f, \theta)+A_m^{\mathrm{r}}(n,f, \theta),
% \end{equation}
% where $A_m^{\mathrm{d}}(t,k, \theta)$ and $A_m^{\mathrm{r}}(t,k, \theta)$ represent the direct-path and reflection-path parts, respectively. 
\subsection{Learning Target}
The direct-path relative transfer function (DP-RTF) encodes the DP-IPD and DP-ILD through its phase and amplitude, defined as:
\begin{equation}
   B_m(f,\theta_k)={A_m(f, \theta_k)} / {A_r(f, \theta_k)},
\end{equation}
where $r$ is the index of one selected reference channel. The DP-IPD, i.e. the phase part $\angle B_m(f,\theta_k)$ is adopted as the learning target. The mean point of the DP-IPD manifold is used as the non-source target\cite{WYBipdnet}. To facilitate optimization with real-valued networks, the learning target is set as concatenating the real and imaginary parts of the complex valued DP-IPD across frequencies, for one microphone pair, defined as:
\begin{equation}
\begin{aligned}
\mathbf{q}(\theta)= 
 [&\cos \angle B_m(1, \theta), \ldots, \cos \angle B_m(F, \theta), \\
& \sin \angle B_m(1, \theta), \ldots, \sin \angle B_m(F, \theta)]^{\top} \in \mathbb{R}^{2 F},
\end{aligned}
\end{equation}
where $^\top$ denotes vector transpose.
The source permutation problem for training is addressed with frame-level permutation invariant training (PIT). During inference, the mean squared errors (MSEs) between the estimated DP-IPD vector and the theoretical DP-IPD vector of candidate directions are computed. The candidate direction with the smallest MSE is selected as the localization result. Candidate directions are uniformly sampled over the entire localization space.
\label{sec:method}
% Assuming there are multiple moving sound sources in a closed environment with noise and  reverberation. sources at a direction $\theta_k$ is observed by two microphones. The multichannel signals in the STFT domain can be defined as:
% \begin{equation}
% X_m(t, f)=\sum_{k=1}^K A_m\left(t, f, \theta_k\right) S_k(t, f)+V_m(t, f),
% \end{equation}
% where $m \in[1, 2]$, $t \in[1, T]$,$f \in[1, F]$ and $k \in[1, K]$ denote the indices of microphone, time frame, frequency and sound sources, respectively. $X_m(t, f)$, $S(t, f)$ and $N_m(t, f)$ represent the STFT coefficients of microphone, source and noise signal, respectively. $A_m\left(t, f, \theta_k\right)$ is the time-varying/dependent room transfer function because of the sound sources are moving. $\theta_k \in[0,\pi]$  represents the azimuth of the $k$-th source.

\subsection{Network Architecture}
This section presents the proposed IPDnet2. The network architecture is illustrated in Fig.\ref{fig:overview}(a). The network takes as input the concatenated real and imaginary parts of the STFT coefficients of the multichannel microphone signals and estimates the DP-IPDs for the $K$ sources. It adopts the previously proposed oSpatialNet \cite{10570301} as the backbone. Specifically, IPDnet2 consists of interleaved cross-band modules and narrow-band modules with a frequency pooling layer and a time pooling layer, a frequency inverse module, and a linear output layer. The input is first processed by a temporal convolutional layer, producing a hidden representation with dimensions $F \times T \times H$. The hidden tensor is then fed into $L$ cascaded cross-band and narrow-band modules (oSpatialNet layers) with a frequency pooling layer and a time pooling layer. The frequency and time pooling operations are applied once to compress the frequency and time dimensions to $F^{\prime}$ and $T^{\prime}$, respectively. Here, $F^{\prime} = F/N$, $N$ is the predefined frequency compression ratio, and $T^{\prime}$ denotes the temporal dimension of the localization outputs, which determines the temporal resolution of localization. Based on some preliminary experiments, the frequency pooling is always applied after the first cross-band module. Time pooling is typically applied after a narrow-band module, but depending on the setup, it can be placed after different narrow-band modules. The frequency inverse module then restores the frequency dimension of the resulting tensor from $F^{\prime}$ to $F$ to recover the original frequency resolution of DP-IPD. The network outputs DP-IPDs track-wisely, each track represents a source and contains $M-1$ (microphone pairs of) estimated DP-IPD vectors. Finally, a linear layer separates the microphone pairs and sources.

\textbf{Narrow-band module: }  
In narrow-band, rich spatial cues can be used for SSL, which are largely leveraged in conventional methods. 
For example, localization features are extracted by narrow-band channel identification\cite{7533416}, coherence test\cite{Mohan2008LocalizationOM}, and direct-path dominance test\cite{Nadiri2014LocalizationOM}. The narrow-band module focuses on the time dimension to learn these information. Considering that DP-IPD varies with time for moving sources, the narrow-band layers are responsible for modeling this temporal evolution of DP-IPD as well. It processes each frequency independently, while all frequencies share the same network parameters. The input is a temporal sequence of a single frequency, defined as: $H^{\text{narrow}}(f) = (\mathbf{h}(1,f), \dots,\mathbf{h}(t,f), \dots , \mathbf{h}(T,f)),$
the superscript $^{\text{narrow}}$ designates the narrow-band module, while the input vector $\mathbf{h}(t,f)$ corresponds to the output produced by the previous cross-band module.
As shown in Fig.\ref{fig:overview}(b), the narrow-band module consists of two Mamba layers. Mamba \cite{Gu2023MambaLS} is a recently proposed architecture based on structured state space sequence models, which has demonstrated high efficiency in modeling both short-term and long-term dependencies in sequential data. 

\textbf{Cross-band module: } 
Similarly to the full-band modules presented in \cite{WYBIS23, WYBipdnet}, the cross-band module aims to learn the full-band correlation of spatial features, such as the linear dependence of DP-IPD to frequency.  Each time frame is processed independently, and all frames share the same network parameters. The architecture of the cross-band module is shown in Fig. \ref{fig:overview} (c), which consists of three cascaded layers: a frequency convolutional layer (F-GConv1d), a across-frequency linear layers (F-Linear), and another frequency convolutional layer. The frequency convolutional layer takes as input a sequence along the frequency axis with a single time frame, defined as: $H^{\text{fconv}}(t) = (\mathbf{h}(t,1), \dots,\mathbf{h}(t,f), \dots , \mathbf{h}(t,F)),$
where the superscript $^{\text{fconv}}$ denotes the frequency convolutional layer. This layer captures local dependencies among neighboring frequency bins. The subsequent across-frequency linear layer jointly processes all frequencies for each hidden dimension, enabling the modeling of full-band dependencies. 

\textbf{Frequency and time pooling:}
A frequency average pooling layer with a pooling factor of $N$ (the predefined frequency compression ratio) is used after the first cross-band module, which linearly compresses the frequency dimension from $F$ to $F^{\prime}$. Afterward, data will be processed with $F^{\prime}$ frequencies and a lower computational cost. 
On the one hand, the representation of DP-IPD requires a very low frequency resolution, as DP-IPD of different frequencies corresponds to one unique TDOA, which means the frequency resolution can be reduced very low. On the other hand, the estimation of DP-IPD is highly based on the time-frequency analysis of spectral and spatial information, which requires sufficient frequency resolution.  

A temporal average pooling layer is applied after one narrow-band module to compress the time dimension to match that of the localization outputs. After time pooling, the subsequent narrow-band layers process time sequences with a lower temporal resolution and a lower computational cost. 
The estimation of DP-IPD relies on the learning of narrow-band spatial information, such as the narrow-band convolution model \cite{7533416}, which requires sufficient temporal resolution. After time pooling, the temporal resolution may not be sufficient for learning certain information, but it would be also sufficient for learning other information, such as the temporal evolution of DP-IPD.   

Overall, careful experimental analysis should be performed to choose a suitable frequency compression ratio and the placement of time pooling.
% Although time pooling is typically applied after a narrow-band module, depending on the specific setup, it may also be placed after other narrow-band modules.

\textbf{Frequency inverse module: } 
The frequency inverse module aims to restore the frequency dimension from $F^{\prime}$ to $F$ to recover the original frequency resolution of the DP-IPD.
As illustrated in Fig.~\ref{fig:overview} (d), it independently restores the compressed $F/N$ sub-bands. Specifically, each compressed frequency band is expanded by a factor of $N$ using a linear layer, and the resulting features are concatenated along the frequency axis to form the full-band output. A $\tanh$ activation function is used after the linear layer.
% \begin{equation}
%     H^f(t) = (\mathbf{h}^\textit{f}(t,1), \dots,\mathbf{h}^\textit{f}(t,k), \dots , \mathbf{h}^\textit{f}(t,K)),
% \end{equation}
% \subsection{Frame-level PIT}
% The source permutation problem in the training stage is solved through the frame-level permutation invariant training (frame-level PIT). For a given frame $t$, $P(t)$ denotes the set of estimated DP-IPDs aligned with the DP-IPD targets. $\alpha \in P(t)$ represents one frame-level permutation. The frame-level PIT loss can be written as follows:
% \begin{equation}
% \mathcal{L}^{\mathrm{PIT}}=\frac{1}{T} \sum_t^T \min _{\alpha \in \operatorname{P}(t)} l_{\alpha, t},
% \end{equation}
% \begin{equation}
% l_{\alpha, t}=\frac{1}{N} \sum_n^N \operatorname{MSE}\left(\boldsymbol{P}_{\alpha, n c t}^*, \hat{\boldsymbol{P}}_{n c t}\right),
% \end{equation}
% \subsection{DP-SNR}
% To define whether a source exists in the current frame, we propose a source flag based on the direct path signal-to-noise ratio (DP-SNR).

% Below is an example of how to insert images. Delete the ``\vspace'' line,
% uncomment the preceding line ``\centerline...'' and replace ``imageX.ps''
% with a suitable PostScript file name.
% -------------------------------------------------------------------------
\section{Experiments}
\subsection{Dataset}
The experiments are conducted on the RealMAN dataset\cite{RealMAN2024}, which is a real-recorded, annotated microphone array dataset for SSL and SE tasks. RealMAN provides 32-channel recordings. In our experiments, channels 0, 1, 3, 5, and 7 are used, which forms by a 4-element planar circular microphone array with a center microphone and a 3 cm radius.
Multi-source microphone signals are generated by mixing two single-source signals from RealMAN. For training, single-source signals are randomly sampled from the dataset and mixed with four overlap strategies from \cite{nbc2} (Head-Tail, small, Start-or-End, and Full), with each strategy applied in equal proportion. Randomly selected real-recorded noise is added to speech signals with a signal-to-noise ratio (SNR) randomly selected from -5 dB to 15 dB. 
% Noise signals are randomly selected from the noise set without matching the source signal scenario. 
For validation and test data, to evaluate the localization performance of different methods in real-world scenarios, speech signals of different speakers recorded in the same recording scene are randomly selected and mixed with an overlap rate uniformly sampled from [0, 1]. Real-recorded noise from the same/similar scene (following the principle presented in \cite{RealMAN2024}) is added directly without volume scaling. 
The signals for the training, validation, and test sets are taken from the corresponding subsets of the RealMAN dataset.

\subsection{Configurations and Comparison methods}
\textbf{Configurations: }The window length of STFT is 512 samples (32 ms) with a frame shift of 320 samples (20 ms). The length of each audio clip is 4 s. The real and imaginary parts of the STFT coefficients are concatenated as the input to the network. The same normalization method for online SSL of \cite{WYBipdnet} is performed on the network input. The proposed model outputs a localization result every 5 STFT frames (100 ms). MSE is used as the loss function. The AdamW optimizer \cite{Loshchilov2017DecoupledWD} with an initial learning rate of 0.0005 is used in the training stage, and the learning rate is exponentially decayed with a decaying factor of 0.975. The batch size is set to 16. The model is trained for almost 80 epochs. $K$ is set to 2. The threshold of 0.4 is set to the estimated spatial spectrum to identify the presence of a source. Performance is only evaluated on voice-active periods. The resolution of candidate azimuths is 1$^{\circ}$. The azimuth estimation error is computed as the difference of estimated and true. Tolerance is set to 5$^{\circ}$, which means that the source is considered to be successfully localized if the azimuth estimation error is not larger than 5$^{\circ}$. Evaluation metrics include miss detection rate (MDR), false alarm rate (FAR) and mean absolute error (MAE). MDR and FAR represent the proportions of frames where the source is active but not successfully localized, and where a source is detected but not active, respectively. MAE represents the absolute angle estimation error of all successfully localized sources and time frames.  Code is available from this link \footnote{https://github.com/Audio-WestlakeU/FN-SSL}.

\textbf{Comparison methods: }
The following advanced approaches are compared with the proposed method:
\textbf{(1) IcoCNN} is an extended version of \cite{DiazGuerra2022DirectionOA} for multiple sources localization, it uses the icosahedral CNN to capture the localization cues from the SRP-PHAT spatial spectrum, while permutation-invariant training is applied for estimating multiple source locations \cite{diaz2023permutation}.
\textbf{(2) SRP-DNN} is a casual CRNN network which uses iterative source detection and localization methods to get multiple localization results\cite{Yang2022SRPDNNLD}.
\textbf{(3) IPDnet} employs full-band BLSTM and narrow-band LSTM to estimate DP-IPDs of multiple sources track-wisely, and subsequently maps the DP-IPD to the source location\cite{WYBipdnet}.
\textbf{(4) TF-Mamba} \cite{Xiao2024TFMambaAT} replaces the LSTM in IPDnet with Mamba to improve the localization performance which was originally proposed for single source localization. We use its Mamba-based backbone and estimate the track-wise IPDs for multiple speakers, as is done in IPDnet and in this work. 
\subsection{Experimental results}
\textbf{The results of the comparison experiments} are presented in Table \ref{table:compare}. Three versions of IPDnet2—tiny, small, and large are proposed for different application scenarios. The hidden dimension $H$ is set to 48, 96, and 256 for the tiny, small, and large models, respectively. All three versions use eight oSpatialNet layers and apply frequency pooling after the first cross-band module with a frequency compression ratio of $N=16$. For time compression, the tiny and small models apply time pooling after the first narrow-band module, whereas the large model applies after the last narrow-band module. All methods except IcoDOA use raw microphone signals and perform substantially better than IcoDOA, which relies on the noisy SRP-PHAT spectrum. Direct processing of microphone signals is more effective in suppressing noise and reverberation by leveraging the properties of original noise and reverberation presented in the microphone signals. The full-band and narrow-band fusion networks outperform other methods, demonstrating that this architecture can efficiently exploit the temporal evolution of narrow-band spatial information as well as the cross-band correlation of localization cues. IPDnet2-Small achieves comparable localization performance to IPDnet while requiring less than 2\% of its computational cost, which demonstrates the effectiveness of the proposed frequency and time pooling mechanism. Additionally, by scaling up the model size, IPDnet2-Large  achieves SOTA localization performance, still with relatively low computational overhead. 
% This indicates that the proposed method achieves superior performance in terms of both localization performance and computational complexity.
\begin{table}[]

    \centering
    \caption{Comparison results. F-N denotes whether it is a full-band and narrow-band fusion network.}
        \vspace{0.5em}
    \footnotesize
    \renewcommand\arraystretch{1.05}
    \tabcolsep0.04in
    \centering

    \begin{tabular}{clccccc}
\toprule
\multirow{2}{*}{Methods} & \multirow{2}{*}{F-N} & FLOPs        & \#Params.    & \multicolumn{3}{c}{Tolerance = 5°}          \\
                         &                      & {[}G/s{]}    & {[}M{]}      & MDR{[}\%{]}   & FAR{[}\%{]}  & MAE{[}°{]}   \\ \midrule
IcoDOA\cite{diaz2023permutation}                   & $\times$                    & 100.7        & 4.7          & 44.6          & 25.5         & 2.3          \\
SRP-DNN\cite{Yang2022SRPDNNLD}                  & $\times$                   & 7.7          & 0.8          & 18.6          & 10.8         & 1.6          \\
IPDnet\cite{WYBipdnet}                   & $\checkmark$                    & 54.4         & 2.1          & 13.8          & 10.0         & 1.4          \\
TF-Mamba\cite{Xiao2024TFMambaAT}                 & $\checkmark$                    & 36.6         & 1.6          & 11.8          & 9.4          & 1.4          \\ \midrule
\rowcolor{gray!20}
IPDnet2-Tiny                     & $\checkmark$                    & \textbf{0.2} & \textbf{0.5} & 17.4          & 11.8         & 1.6          \\
\rowcolor{gray!20}
IPDnet2-Small                    & $\checkmark$                    & 0.9          & 1.3          & 13.5          & 9.5          & 1.4          \\
\rowcolor{gray!20}
IPDnet2-Large                     & $\checkmark$                    & 13.9         & 7.8          & \textbf{10.5} & \textbf{6.6} & \textbf{1.2} \\ \bottomrule
\end{tabular}

    \label{table:compare}
\end{table}

\begin{figure}[t]
    \centering
    \includegraphics[width=1\linewidth]{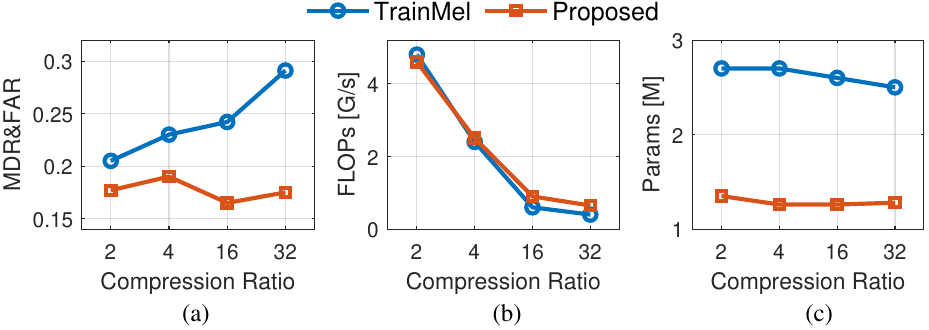}
    \caption{(a) SSL performance, (b) FLOPs, and (c) model size as a function of frequency compression ratio.}
    \label{fig:fig1_v2}
\end{figure}

\begin{table}[t]
    \caption{Ablation experiments.}
    \vspace{0.5em}
\footnotesize
    \centering
    \renewcommand\arraystretch{1.1}
\tabcolsep0.02in % 可以稍微调大些
   \begin{tabular}{clcccccc}
\toprule
\multicolumn{2}{c}{\multirow{2}{*}{Methods}} & FLOPs & \#Params. & MDR & FAR & MAE \\
\multicolumn{2}{c}{} & {[}G/s{]} & {[}M{]} & {[\%]} & {[\%]} & {[°]} \\ \midrule

\rowcolor{gray!20}
\multicolumn{2}{l}{IPDnet2-Small (prop.)}   & 0.9 & 1.3 & \textbf{13.5} & \textbf{9.5} & \textbf{1.4} \\
               & \hspace{0.4em}freq. pooling before first cross-band  & 0.7 & 1.2 & 24.3 & 13.5 & 1.7 \\
               & \hspace{0.4em}time. pooling before first cross-band   & 0.6 & 1.3 & 18.1 & 12.4 & 1.6 \\
               & \hspace{0.4em}with 1 SpatialNet layer         & 0.6 & 0.2 & 18.3 & 11.3 & 1.6 \\ 
\bottomrule
\end{tabular} 

    \label{tab:ablation}
\end{table}
\textbf{Frequency compression comparisons }are shown in Fig.\ref{fig:fig1_v2}. Under the same IPDnet2 architecture, we compared different compression mechanisms. MDR\&FAR is defined as the square root of the sum of squared MDR and squared FAR, which reflects the overall performance of SSL. TrainMel\cite{trainmel} is an advanced frequency compression mechanism, originally designed to reduce the computational cost for speech enhancement. It compresses the input frequency dimension using grouped fully connected layers. As illustrated in Fig.\ref{fig:fig1_v2} (a), TrainMel exhibits substantial degradation in localization performance as the compression ratio increases from $2$ to $32$. The possible reason is that TrainMel is designed to conduct frequency compression directly to the input signal, which is less effective for SSL, although a more complex learnable frequency compression network is employed. By contrast, as proposed in this work, after learning full-band information with one cross-band module, and then applying a simple frequency average pooling would capture sufficient full-band information for SSL. Overall, up to a compression ratio of $16$, the proposed frequency compression mechanism achieves promising localization performance, significant reduction of computation complexity. 
% In contrast, the proposed method incurs only limited performance loss. Under the same compression settings and with comparable computational cost (FLOPs
% % \footnote{FLOPs (in Giga per second, G/s) are computed on 4 s utterances and divided by 4, using the official PyTorch tool (torch.utils.flop\_counter.FlopCounterMode).}
% ), it markedly outperforms TrainMel. Notably, it achieves the best performance at a $16$ frequency compression, indicating that frequency compression in SSL can be performed in a simple way within the linear frequency domain. For example, average pooling, as used in this work, is sufficient without incurring information loss. The advantages of the proposed method are as follows: i) It performs compression after frequency feature extraction, whereas TrainMel directly compresses the raw inputs, leading to information loss along the frequency axis. ii) the proposed method achieves compression without additional trainable parameters. As the Fig.\ref{fig:fig1_v2} (c) shown, TrainMel relies on grouped linear layers that introduce extra parameters.

% \textbf{Comparing with other methods.}
% \begin{table}[]
%     \centering
%     \begin{tabular}{c|c}
%          &  \\
%          & 
%     \end{tabular}
%     \caption{Caption}
%     \label{tab:placeholder}
% \end{table}

\textbf{The ablation results} are presented in Table \ref{tab:ablation}. 
% where FP-Front and TP-Front denote placing frequency pooling and time pooling, respectively, before the first cross-band module. 
Experiments are conducted with the IPDnet2-Small model, in which frequency pooling and time pooling are applied after the first cross-band and narrow-band layers respectively. When putting frequency pooling or time pooling before the cross-band module, the localization performance will be largely degraded, which indicates at least one layer of cross-band and narrow-band processing are very important for learning useful SSL information in the high resolution of frequency and time dimensions. Most of the computations in IPDnet2-Small are concentrated in the first oSpatialNet layer ($\approx$2/3 of total cost) due to compression. We compare with when using only this 1 layer, it can be seen that the subsequent 7 layers after pooling are very important as well, and useful SSL information can be learned in the low-resolution domain.

\section{Conclusions}
In this work, we extend IPDnet to IPDnet2 to improve localization accuracy and reduce computational complexity. oSpatialNet is adopted as backbone network to enhance the capability of spatial feature extraction and the scalability of model size. In addition, we propose an effective frequency and time pooling mechanism that respectively compresses frequency and time, thereby substantially reducing computational cost. As a result, IPDnet2 matches IPDnet’s localization performance with less than 2\% of its computation cost, and achieves state-of-the-art SSL performance with a larger version of the model. 
% In future work, we will focus on training the variable-array model of IPDnet2 with real-recorded data to provide a low-cost solution to the scarcity of array-specific real-world data.

\bibliographystyle{IEEEbib}
\bibliography{strings,refs}

\end{document}